\pdfoutput=1

\documentclass[11pt]{article}

\usepackage[preprint]{acl}

\usepackage{times}
\usepackage{latexsym}

\usepackage[T1]{fontenc}

\usepackage[utf8]{inputenc}

\usepackage{microtype}

\usepackage{inconsolata}

\usepackage{graphicx}

\usepackage{float}
\usepackage{makecell}
\usepackage{pifont}
\usepackage{tikz}
\usepackage[edges]{forest}
\usepackage{multirow}
\usepackage{booktabs}
\usepackage{amsmath, amsfonts}

\title{Beyond A Single AI Cluster: A Survey of Decentralized LLM Training}

\author{
Haotian Dong\textsuperscript{1}\thanks{~~Equal Contribution.}, 
Jingyan Jiang\textsuperscript{2}\footnotemark[1],
Rongwei Lu\textsuperscript{1},
Jiajun Luo\textsuperscript{1}, \\
\textbf{
Jiajun Song\textsuperscript{1},
Bowen Li\textsuperscript{1}, 
Ying Shen\textsuperscript{3}\thanks{~~Corresponding Author.},
Zhi Wang\textsuperscript{1}\footnotemark[2]
} \\
\textsuperscript{1}Shenzhen International Graduate School, Tsinghua University \\
\textsuperscript{2}Shenzhen Technology University 
\textsuperscript{3}China Central Depository \& Clearing Co., Ltd.\\
\texttt{donght24@mails.tsinghua.edu.cn},
\texttt{jiangjingyan@sztu.edu.cn}\\
\texttt{wangzhi@sz.tsinghua.edu.cn}
}

\begin{document}
\maketitle
\begin{abstract}
The emergence of large language models (LLMs) has revolutionized AI development, yet their resource demands beyond a single cluster or even datacenter, limiting accessibility to well-resourced organizations.
Decentralized training has emerged as a promising paradigm to leverage dispersed resources across clusters, datacenters and even regions, offering the potential to democratize LLM development for broader communities.
As the first comprehensive exploration of this emerging field, we present decentralized LLM training as a resource-driven paradigm and categorize existing efforts into community-driven and organizational approaches.
We further clarify this through: (1) a comparison with related paradigms, (2) characterization of decentralized resources, and (3) a taxonomy of recent advancements. We also provide up-to-date case studies and outline future directions to advance research in decentralized LLM training.
\end{abstract}

\section{Introduction}

The rapid advancement of LLMs has yielded remarkable progress across a wide range of domains~\cite{DeepSeekAI2024DeepSeekV3TR}. With model scales expanding from GPT-3's \cite{Brown2020LanguageMA} 175 billion to DeepSeek-R1's \cite{DeepSeekR1} 660 billion parameters, the computing resource demands of training LLMs have increased dramatically \cite{Jiang2024MegaScaleSL}. 

However, this exponential growth in computational requirements poses significant challenges. For individual researchers and small laboratories with limited resources, the demands are particularly prohibitive.
Even for well-resourced organizations, confining LLM training within a single AI cluster faces challenges like: geographically distributed services required for latency optimization~\cite{mcmahan2017communication}, inherent hardware bottlenecks limiting single-cluster scalability~\cite{Athlur2022VarunaSL,Dubey2024TheL3}, economic patterns requiring placement adaptation \cite{Liu2023placement}, etc.

These challenges underscore the necessity for innovative resource management approaches to enhance the accessibility of LLM training. One such approach is \emph{decentralized LLM training}, a distributed paradigm that leverages decentralized resources at varying scales to achieve greater scalability and cost-efficiency.

Training LLMs with decentralized resources faces inherent challenges across different scenarios. 
For individual researchers and communities collaborating in decentralized environments, key challenges include dynamic resource availability, limited bandwidth in wide area networks (WANs), and heterogeneous computing capabilities \cite{Borzunov2022PetalsCI,Yuan2022DecentralizedTO,Yang2023HolmesTD}.
For well-resourced organizations managing multiple clusters or datacenters, it is essential to consider not only communication efficiency but also energy consumption minimization and cross-datacenter workload scheduling coordination~\cite{Park2024CarbonAwareAF,Choudhury2024MASTGS}.
Furthermore, the inherent complexity of the LLM training process exacerbates these challenges. Modern LLM training relies on a hybrid of parallelization strategies to efficiently coordinate computing resources~\cite{Narayanan2021EfficientLL}, which significantly amplifies the difficulties when operating within a decentralized infrastructure.

\begin{table*}[!ht]
\centering

\resizebox{\linewidth}{!}{%
\begin{tabular}{lccccc}
\toprule
\textbf{Surveys} & \textbf{LLM-Focused} & \textbf{Resource-Driven} & \textbf{Cross-Regional} & \textbf{Paradigms}\\
\midrule
\cite{Duan2024EfficientTO} & \checkmark & \checkmark & &
\makecell[c]{Efficient LLM Training, Centralized Infrastructures}\\ 
\cite{Khan2023DecentralizedML} &  &  & \checkmark & 
\makecell[c]{Decentralized Machine Learning, Geo-Distributed Machine Learning} \\
\cite{Woisetschlger:FLFM}  & \checkmark  & & \checkmark & 
\makecell[c]{Federated Learning, Efficient Foundation Model Training} \\
\textbf{Ours} & \checkmark & \checkmark & \checkmark & 
\makecell[c]{Decentralized LLM Training, Decentralized Infrastructures} \\
\bottomrule
\end{tabular}%
}

\caption{Comparison with related surveys. These paradigms overlap in terms of scenarios and optimization targets. To illustrate decentralized LLM training and resources, comparison and analysis are presented in \S \ref{sec:background} and \S \ref{sec:resources}.}
\label{tab:notations}
\end{table*}

\begin{figure*}[ht]
    \centering
    \includegraphics[width=1\linewidth]{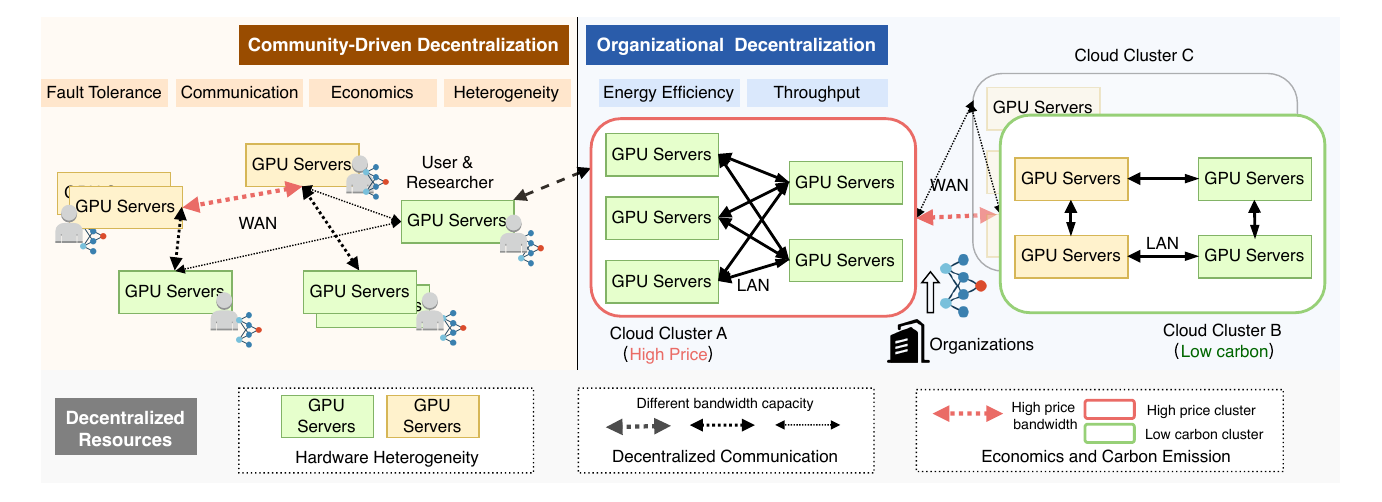}
    \caption{Decentralized LLM training paradigms of both community-driven and organizational  decentralization. \textbf{Community-driven decentralization} utilizes pooling resources from communities with independent entities. These pooled resources can include local servers from researchers, instances from cloud services or volunteer communities, etc. \textbf{Organizational decentralization} involves consolidating resources from multiple clusters or even multiple datacenters managed by well-resourced organizations (e.g., technology giants and governments).}
    \label{fig:framework}
\end{figure*}

This survey systematically investigates challenges and solutions for decentralized LLM training.
Compared to prior surveys on other distributed paradigms, our paper centers around LLM training with decentralized resources, as shown in Table~\ref{tab:notations}. 
We aim to characterize the utilization of decentralized resources and analyze optimization methods in decentralized LLM training, thereby exploring potential research opportunities.
Figure \ref{fig:framework} illustrates the utilization paradigms of resources in decentralized LLM training and optimization objectives we categorize in this paper.
To the best of our knowledge, our survey is the \textit{first} to study recent advances in decentralized LLM training. Our work contains three distinct contributions:

\begin{itemize}
    \item \textbf{A resource-driven position of decentralized LLM training}.
    We position decentralized LLM training as a resource-driven paradigm by comparing with related paradigms and examining the characteristics of decentralized resources. 
    \item \textbf{A novel taxonomy on decentralization paradigms and optimization objectives of decentralized LLM training}. 
    We classify the utilization of decentralized resources into two paradigms: community-driven and organizational.
    Then we taxonomize recent studies based on optimization objectives and review related methodologies.
    \item \textbf{Up-to-date case studies and prospective future research directions}. 
    We compare two LLMs trained under different decentralized paradigms: one leveraging organizational resources and the other utilizing scattered, fragmented resources. We also propose potential directions for future research.
    
\end{itemize}

\section{Background}
\label{sec:background}

The concept of decentralized LLM training intersects with several distributed machine learning paradigms, including \textit{Geographically-Distributed Machine Learning (Geo-ML)}, \textit{Federated Learning (FL)}, \textit{Decentralized Machine Learning (De-ML)}, and \textit{Efficient LLM Training}. We review these paradigms to highlight the distinctive characteristics of decentralized LLM training as a resource-driven paradigm.

\subsection{Geo-ML and FL}

Geo-ML and FL both tackle challenges from data decentralization, yet each approach has its distinct focus. Geo-ML mainly tackles service latency and regulatory requirements by optimizing training processes across datacenters, leveraging hierarchical network topologies (e.g., high-bandwidth local area networks (LANs) within datacenters and limited-bandwidth WANs between datacenters) to utilize decentralized data and computational resources efficiently \cite{Hsieh2017GaiaGM}. In contrast, FL focuses more on privacy protection \cite{mcmahan2017communication}, employing either centralized parameter-server or decentralized methods~\cite{Lalitha2019PeertopeerFL,Xing2021FederatedLO}, with resources from decentralized clusters or edge devices.

\subsection{De-ML}

De-ML has a dual meaning in terms of decentralization. From the distributed strategy perspective, it mitigates communication bottlenecks inherent in parameter server architectures through peer-to-peer networking~\cite{Hegeds2019GossipLA,WarnatHerresthal2021SwarmLF}. From the resource utilization perspective, it offers a cost-effective and flexible strategy that maximizes the utilization of geographically dispersed computing resources \cite{Yuan2022DecentralizedTO}. While both De-ML and Geo-ML involve training architectures, they differ fundamentally in their approach---the former adopts peer-to-peer topology where nodes communicate directly, while the latter relies on hierarchical centralized architecture both across and within datacenters.

\subsection{Efficient LLM Training}

Currently, LLM training primarily employs distributed methods with centralized resources. However, efficient resource utilization remains challenging even within a single cluster. Efficient LLM training methods (e.g. 3D parallelism \cite{Narayanan2021EfficientLL}, mixed-precision training \cite{Micikevicius2017MixedPT}, gradient compression \cite{dagcTmc}, etc.) have been proposed to address scalability, efficiency, and reliability challenges \cite{Duan2024EfficientTO}. These approaches are also essential for decentralized LLM training, where resource constraints are more severe and complex.

\subsection{Decentralized LLM Training}

\begin{center}
\fcolorbox{yellow!10}{yellow!10}{\parbox{.97\linewidth}
{\textbf{Position.} \textit{We position decentralized LLM training as the convergence of decentralized ML and efficient LLM training, primarily driven by resources distribution akin to Geo-ML.}}}
\end{center}

Decentralized LLM training leverages distributed resources at varying scales to meet the substantial computational demands. 
Resource-limited communities rely on fragmented, globally contributed resources to meet LLM training demands. For well-resourced organizations, the resource demands for training ultra-scale LLMs necessitate the combination of consolidated resources from multiple clusters or datacenters to enable relatively decentralized training \cite{Alibaba_HPN,Dubey2024TheL3}. 
Based on these resource utilization paradigms, we categorize decentralized LLM training into: \textit{community-driven decentralization} and \textit{organizational decentralization}. These two decentralization paradigms are depicted in Figure \ref{fig:framework}.

The fundamental distinction between utilizing decentralized resources for training LLMs and general models lies in the demand of more sophisticated parallelization strategies. In LLM training, the substantial model parameter scale coupled with extensive activation values generated by auto-regressive Transformer architecture (e.g., GPT-3's activation memory during training exceeds its model parameters by more than 5$\times$ \cite{Narayanan2021EfficientLL}), rendering simple Data Parallelism (DP) inadequate. This necessitates the adoption of strategies like Tensor Parallelism (TP) and Pipeline Parallelism (PP) to effectively distribute activation memory across GPUs, typically forming a comprehensive 3D parallelism.
Decentralized LLM training leverage a broader range of resources compared to traditional distributed LLM training, yet it still employ similar parallel strategies.
DP and PP are currently the primary parallel strategies\footnotemark[1] for training LLMs with decentralized resources \cite{Yuan2022DecentralizedTO,Ryabinin2023SWARMPT}. Due to intensive communication requirements, TP is more constrained within both LANs and WANs.

\footnotetext[1]{More details about parallel strategies of distributed training are presented in Appendix \ref{sec:appendix_a}}

\section{Decentralized Resources}
\label{sec:resources}
Training LLMs with decentralized resources efficiently is challenging due to resource constraints in communication, computational heterogeneity, and cost disparities. We present characteristics of decentralized resources to illustrate these constraints in this section.

\subsection{Communication Constraint}

In distributed training, computational devices need to communicate frequently. However, decentralized resources have more limited communication capabilities compared to centralized ones.

Community-driven decentralized resources, combining with local servers and cloud instances, communicate through LANs or even WANs during model training with bandwidths under 10 Gb/s~\cite{azure}.
In contrast, organization-driven decentralized training can reach over 500 GB/s of bandwidth within a server by NVLink, or over 200 Gb/s within a cluster using InfiniBand.
However, when the scale of resources expands across multiple geographically distributed clusters or datacenters, the bandwidth can drop to only hundreds of Mb/s \cite{xiang2022nebula}.
To alleviate communication bottlenecks, strategies such as gradient compression \cite{dagcTmc,Lu2025FedHTSH}  and delayed aggregation \cite{DGA,Lu2025DeCoSGDJO} are often employed during distributed training.

\subsection{Hardware Heterogeneity}

In distributed training, computational devices are typically configured with identical specifications (e.g., memory capacity and FLOPS\footnotemark[2]) to prevent slower ones from becoming bottlenecks \cite{shen2024efficient}. 

However, in decentralized paradigms, when aggregating resources across multiple nodes, clusters, or datacenters, the heterogeneity of resources becomes significant. This is particularly evident in communication, computation, and hardware architecture \cite{Yuan2022DecentralizedTO,Yang2023HolmesTD}. For instance, in geographically distributed clusters, computational devices like GPUs and NPUs with different architectures may coexist, and the communication bandwidth between clusters varies \cite{xiang2022nebula}. Due to the fast-paced evolution of computing devices, hardware heterogeneity is almost inevitable \cite{Tang2023FusionAIDT}.

\footnotetext[2]{FLOPS (Floating-Point Operations Per Second) measures the number of floating-point arithmetic operations (e.g., addition, subtraction, multiplication, division) that a computational unit can perform in one second.}

\subsection{Cost Volatility}

Training ultra-scale LLMs typically requires tens of thousands of GPUs running for thousands of hours, resulting in significant computational costs and energy consumption \cite{Dubey2024TheL3}. 

However, reducing costs with decentralized resources is often more challenging than centralized setups.
In community-driven decentralization, fragmented resources such as preemptive cloud instances are relatively cheaper, but their prices are highly volatile \cite{Lee2017DeepSpotCloudLC}. Additionally, resource instability can lead to re-computation and inter-node waiting, prolonging training process and increasing overall costs.
For organizational decentralization, 
accurately modeling energy consumption and optimizing resource scheduling become increasingly complex, owing to the demand for massive resources and the inherently cross-datacenter nature of this paradigm \cite{Faiz2023LLMCarbonMT,Choudhury2024MASTGS}.


\section{Community-Driven Decentralization}

Training LLMs under community-driven paradigm often encounters challenges, including hardware heterogeneity, limited communication bandwidth, resource instability, and economic volatility. This chapter focuses on these challenges and provides an in-depth discussion of related research efforts.
We summarize and compare selected studies in Table \ref{tab:com_dec}. A more comprehensive paper list is presented in Figure \ref{fig:papers_community} of Appendix \ref{sec:appendix_b}.

\begin{table*}[!ht]
\centering
\resizebox{\linewidth}{!}{%
\begin{tabular}{lccccccc}
\toprule
\textbf{Papers} & \textbf{Parallel Strategy} & \textbf{Model Type} & \textbf{Resource Scale} & \textbf{Comm.} & \textbf{Hete.} & \textbf{Faul.} & \textbf{Econ.} \\
\midrule
L@H \cite{Ryabinin2020TowardsCT} & Decentralized Mixture-of-Experts & Transformer-XL (1.3B, MoE) & 4 GTX 1080 GPUs & \checkmark & \checkmark & \checkmark &\\
DeDloc \cite{Diskin2021DistributedDL} & DP with extremely large batches & BERT (170M) & 91 devices (RTX 2060, K80, etc.) & \checkmark & \checkmark & \checkmark & \\ 
Moshpit \cite{Ryabinin2021MoshpitSC} & DP with dynamic local groups & ALBERT (18M) & 8 V100 GPUs and 66 GPU instances  & \checkmark & \checkmark & \checkmark &\\
Varuna \cite{Athlur2022VarunaSL} & PP with inner-stage DP   & GPT-2 (200B)  & Low-priority spot VMs with 300 GPUs & \checkmark &  & \checkmark \\
AQ-SGD \cite{Wang2022FinetuningLM} & PP with inner-stage DP & GPT2 (1.5B) & 32 V100 GPU instances & \checkmark & & & \\
DTFM \cite{Yuan2022DecentralizedTO} & PP with inner-stage DP & GPT3 (1.3B) & 64 V100 GPUs of 8 nodes & \checkmark & \checkmark & & \\
SWARM \cite{Ryabinin2023SWARMPT}  & PP with inner-stage DP & Transformer (1.01B) & Spot instances with 400 T4 GPUs & \checkmark & \checkmark & \checkmark & \\
Petals \cite{Borzunov2022PetalsCI} & PP with dynamic stage & BLOOM (176B) & 27 GPUs (A100, RTX3090, A4000, etc.) & \checkmark & \checkmark & \checkmark & \\
FusionAI \cite{Tang2023FusionAIDT} & PP with load-balance scheduling & GPT-3 & 50 RTX 3080 GPUs & \checkmark & & \checkmark & \\
Ravnest \cite{Menon2024RavnestDA} & PP with inner-stage DP & BERT-base (110M) &  10 nodes (A10G, V100, T4) in 4 clusters & \checkmark & \checkmark & \checkmark & \\
StellaTrain \cite{StellaTrain2024} & DP with dynamic batch & GPT-2 (123.6M) & 10 GPUs (V100, RTX4090, etc. )& \checkmark & \checkmark & & \checkmark \\
Holmes \cite{Yang2023HolmesTD} & PP with inner-stage DP and TP & GPT (7.5B) & 64 A100 GPUs of 8 nodes & \checkmark & \checkmark & & \\
Atom \cite{Wu2024ATOMAT} & DP with memory swapping & GPT-3 (13B) & 12 GPUs (V100, RTX 1080Ti, etc.) of 3 nodes &  & \checkmark & \checkmark & \\
Positon \cite{Lu2024PositionET} & PP with layer skipping & Bloom (7B)  & 6 nodes (A40, V100, etc. ) &  &  & \checkmark \\
MLTC \cite{Strati2024MLTW} & PP and DP comparison & OPT (30B) & 85 A100 GPUs & & & & \checkmark \\
DiLoCo \cite{Douillard2023DiLoCoDL}  & DP with large local steps  & Simple Transformer (400M) & 8 A100 nodes (16GPUs each) & \checkmark & & \checkmark & \\
HowCW \cite{Isenko2023HowCW} & DP with target batch size & RoBERTa-XLM (560.1M) & 8 VMs cross continents & & & \checkmark & \checkmark \\
\bottomrule
\end{tabular}
}

\caption{Comparison of related works in community-driven decentralization. The symbol \checkmark indicates whether a paper primarily focuses on a specific optimization objective. Comm: communication efficiency; Hete: network or device heterogeneity; Faul: fault tolerance; Econ: economics. For the model type, the largest language model used in each paper is selected. Papers where language models were not explicitly used are included in Appendix \ref{sec:appendix_b}.}
\label{tab:com_dec}
\end{table*}

\subsection{Communication Optimization}
\label{sec:comm_optimization}

Due to bandwidth limitations of LANs and WANs, it is essential to optimize communication efficiency for decentralized LLM training to alleviate bottleneck. Primary strategies can be divided into optimizations at both temporal and spatial levels, corresponding to: (1) reduce communication frequency; (2) reduce communication intensity.

\paragraph{Reduce Communication Frequency.}
During community-driven decentralization, the fragmented resources used are often distributed across regions or even globally. As a result, the communication process involves local and global levels, in which the global often becomes the bottleneck.  

In DP-only strategy, the primary communication payload comes from gradient transmission.
Gaia~\cite{Hsieh2017GaiaGM} dynamically eliminate insignificant gradients to reduce communication across datacenters.
Co-learning \cite{Xu2018CollaborativeDL} enlarge the number of local epochs dynamically to reduce global synchronization frequency between datacenters.
DeDloc \cite{Diskin2021DistributedDL} adopt large local batches while training to allow peers to communicate less frequently.
DiLoCo \cite{Douillard2023DiLoCoDL} only synchronize globally once after 500 local optimization steps, effectively reducing communication frequency.

In PP strategy, gradients are exchanged within a stage, while both gradients and activations are transmitted across stages.
Varuna \cite{Athlur2022VarunaSL} leverages rule-based policy to adjust PP depth based on the available GPU count to better accommodate bandwidth constraints. 
To allocating tasklets requiring high communication volume to computing units with faster connections, DTFM~\cite{Yuan2022DecentralizedTO} uses a two-level approach: a balanced graph partitioning problem for DP within each stage, and a joint graph matching and traveling salesman problem for the entire PP process.
More dynamically, SWARM \cite{Ryabinin2023SWARMPT} optimizes the PP process by enabling real-time adjustments during each iteration. In this strategy, each PP stage uses multiple candidate devices. When a device outperforms others, it processes inputs from multiple slower predecessors and distributes outputs to multiple slow successors, maximizing bandwidth utilization.

\paragraph{Reduce Communication Intensity.}
Similar to training LLMs with centralized resources, employing techniques such as compression and sparsification for gradients or activations to eliminate insignificant values is an effective approach to reducing whole communication intensity.

For gradients, StellaTrain \cite{StellaTrain2024} leverages gradient sparsity to achieve a 99\% compression rate, significantly reducing communication intensity. 
OpenDiLoCo \cite{jaghouar2024opendiloco} employs mixed precision training \cite{Micikevicius2017MixedPT} with FP16 quantized gradients to reduce communication. 
For activations, SWARM~\cite{Ryabinin2023SWARMPT} and Petals \cite{Borzunov2022PetalsCI} use quantization method to reduce activations. Rather than compressing activation values directly, AQ-SGD \cite{Wang2022FinetuningLM} transmits and compresses sparser activation changes.
With decentralized mixture of experts (DeMoE) structure, Learning@Home \cite{Ryabinin2020TowardsCT} diminishes activations through expert selection to reduce communication payload.

Coordinating the communication topology can balance payload with multiple connections, thereby reduce the intensity of a single session.
Moshpit~\cite{Ryabinin2021MoshpitSC} dynamically forms communication groups to reduce network load during all-reduce. 
Ravnest \cite{Menon2024RavnestDA} parallelizes multiple Ring All-Reduce operations simultaneously to accommodate low bandwidth.

Additionally, optimizing parallel strategies to overlap communication with computation can also accommodate bandwidth limitations in decentralized LLM training. In this condition, PP is often employed \cite{Athlur2022VarunaSL,Yuan2022DecentralizedTO,Ryabinin2023SWARMPT,StellaTrain2024}. We regard it as one method to reduce communication intensity as PP often leads to small-batch communications.

\paragraph{Discussion.}
One of the most significant challenges in decentralized LLM training lies in communication constraints. 
The effectiveness of PP in such settings highlights that developing intelligent communication protocols tailored to LLM-specific features (e.g. 3D parallelism, mixture-of-experts (MoE) and inter-layer parameter sparsity, etc.) is vital for improving the performance and scalability of LLMs in decentralized paradigms.
\subsection{Heterogeneity Awareness}
\label{sec:hete_aware}

Since network bandwidth across regions in WANs varies significantly and computational devices differ in capacity, both network and device heterogeneity exist within decentralized LLM training.

\paragraph{Network Heterogeneity.}
Bandwidth constraints necessitate optimization of communication volume in decentralized LLM training, while network heterogeneity introduces additional challenges in communication scheduling.

To deal with network heterogeneity, BACombo~\cite{Jiang2020BAComboBandwidthAwareDF} uses a bandwidth-aware worker selection strategy, enabling efficient gradient splitting and scheduling across multiple connections. DTFM \cite{Yuan2022DecentralizedTO} employs a communication matrix to model bandwidth differences, and it solves the hierarchical optimization problem to minimize communication cost.
In contrast to static scheduling, DeDLOC \cite{Diskin2021DistributedDL} dynamically adapts its strategy by switching between All-Reduce, parameter servers, and decentralized SGD based on real-time network conditions.
To further optimize communication paths, NETSTOR~\cite{Li2024AcceleratingGM} employs a multi-root adaptive synchronization topology to dynamically allocate tasks based on bandwidth and balancing communication loads with auxiliary paths.
Additionally, hardware characteristics can be integrated into the scheduling process. Holmes \cite{Yang2023HolmesTD} automatically selects heterogeneous network interface cards (NICs, including InfiniBand, RoCE, Ethernet) for diverse distributed strategies (DP, PP and TP) across heterogeneous clusters.

Beyond scheduling, adaptive gradient compression techniques, which dynamically adjust compression levels according to bandwidth, can mitigate straggler issues within heterogeneous networks \cite{Fan2023SelfAdaptiveGQ}.

\paragraph{Device Heterogeneity.}
The varying computational capabilities of decentralized devices significantly influence the strategies for both data-parallel task allocation and model placement.

For data-parallel task allocation, DLion \cite{Hong2020DLionDD} assigns different batch sizes to computational devices in micro-clouds based on their capacities in DP setting. 
SWARM \cite{Ryabinin2023SWARMPT} adaptively merges outputs from predecessors to faster devices and distributes to multiple slower successors within a dynamic PP process.
For model placement,
Learning@home \cite{Ryabinin2020TowardsCT} employs DeMoE paradigm, distributing expert layers to different consumer-grade devices based on memory capability. 
ATOM \cite{Wu2024ATOMAT} performs dynamic model partitioning by jointly considering memory and compute capacity.
Petals~\cite{Borzunov2022PetalsCI} also adopts load balancing, dynamically assigning transformer blocks based on device capability for fine-tuning LLMs on heterogeneous devices.
Ravnest \cite{Menon2024RavnestDA} applies a genetic algorithm to group devices with similar memory and bandwidth, implementing PP within each group and proportionally partitioning model based on capacity to optimize heterogeneous device utilization.

\footnotetext[3]{\url{https://developer.nvidia.com/cuda-toolkit}}
\footnotetext[4]{\url{https://www.hiascend.com/software/cann}}

\paragraph{Discussion. }
{
The effective utilization of decentralized resources is fundamentally constrained by the ability to manage system heterogeneity.
Current frameworks rarely account for cross-architecture collaboration (e.g., NVIDIA GPUs with CUDA\footnotemark[3] and Huawei Ascend GPUs with CANN\footnotemark[4]), which limits their potential.
Furthermore, existing heterogeneity-aware strategies mainly focus on protocol or topology optimization, overlooking critical semantic aspects of LLM training dynamics, such as layer sensitivity and gradient rank \cite{refael2024adarank}. 
Developing semantics-guided, architecture-agnostic frameworks may enhance scalability and resource efficiency in decentralized LLM training.
}

\subsection{Fault Tolerance}
\label{sec:fault_tolerance}

Given the inherent instability of computational resources contributed by communities, node failures and communication disruptions are inevitable in decentralized LLM training. Consequently, fault tolerance becomes a critical requirement to ensure stable and efficient training processes.

A fundamental approach to fault tolerance involves periodically saving model checkpoints and migrating tasks to functioning nodes in the event of failures \cite{Lee2017DeepSpotCloudLC,Athlur2022VarunaSL}. While straightforward, this reactive method is often inefficient due to the overhead of migration and global recomputation. 
To mitigate the impact of single-node failures on global synchronization performance, a hierarchical synchronization strategy has been proposed \cite{Ryabinin2021MoshpitSC,shihang2021breaking} to synchronize locally before global synchronization, which confines recomputation to subgroups, with group size adaptable to failure rates \cite{Diskin2021DistributedDL}.
Waiting for failure nodes recomputation is not always necessary, DiLoCo~\cite{Douillard2023DiLoCoDL} enabling asynchronous local training during communication failures, reducing reliance on global synchronization. Furthermore, if sufficient resources are available, introducing redundant computing replicas to mitigate fault-induced losses can also be a feasible approach \cite{Lu2024PositionET}.

To implement fault-tolerant mechanisms mentioned above, distributed hash tables (DHTs) have emerged as a core technology.
We present how DHTs are leveraged in decentralized LLM training in Appendix \ref{sec:appendix_c}.

\paragraph{Discussion. } Fault tolerance should balance robustness and resource efficiency, especially when employing intricate parallel strategies. For instance, in PP, node failures can lead to complete pipeline stalls. 
Although coarse-grained redundancy and checkpointing mitigate the impact of node failures, they often result in substantial resource inefficiencies. Optimizing this trade-off between performance and resource utilization is crucial for enabling reliable and scalable decentralized training of ultra-large LLMs, ultimately making such training more accessible and reliable.

\subsection{Economics}
\label{sec:economics}

While decentralized resources offer cost advantages compared to dedicated cluster, the volatile pricing of cloud resources, particularly GPU instances, necessitates strategic economic optimization for effective cost control and reduction. 
Prior work shows that 50 commodity RTX 3080 GPUs can deliver throughput comparable to four H100 GPUs, revealing a favorable trade-off between cost and performance ~\cite{Tang2023FusionAIDT}.

One effective cost-saving strategy is the use of low-priority or preemptible instances, which are substantially cheaper than dedicated servers. For instance, Varuna~\cite{Athlur2022VarunaSL} leverages low-priority virtual machines (VMs) that cost approximately 5× less than dedicated GPU servers, without sacrificing training throughput.
Similarly,~\cite{Strati2024MLTW} demonstrates that spot instances, which are 60\%–90\% cheaper than on-demand alternatives, can be effectively used when combined with robust fault-tolerance mechanisms.

Cross-region and multi-cloud resource selection also plays a crucial role in optimizing training economics. DeepSpotCloud~\cite{Lee2017DeepSpotCloudLC} monitors real-time GPU Spot pricing and selects optimal placements to maximize cost-effectiveness. \citet{Isenko2023HowCW} evaluates a hybrid deployment strategy across four continents, finding that using distributed spot instances outperforms centralized DGX-2 or LambdaLabs A10 setups in terms of cost efficiency. StellaTrain~\cite{StellaTrain2024} further explores the hybrid cloud/on-premise setting, reporting a 64.5\% reduction in cloud costs through workload-aware scheduling.

Additionally, several studies propose analytical cost models to guide deployment decisions. \citet{Strati2024MLTW} develops a training cost estimator that highlights the benefits of intra-region communication for improving throughput and minimizing expenses. \citet{Phalak2024TowardsGT} extends this to a multi-cloud, multi-geography scenario, incorporating serverless compute and VM selection into a unified model for performance-cost optimization.

\section{Organizational Decentralization}

In organizational paradigms, besides the optimizations involved in community-driven decentralization, well-resourced organizations consider objectives that extend beyond individual systems to include datacenter-level enhancements, such as energy efficiency, system throughput\footnotemark[5] within and across datacenters.
Representative related papers are compactly summarized in Figure \ref{papers_organization}.

\footnotetext[5] {System throughput refers to the amount of data or tasks a training system can process per unit of time, reflecting its overall processing capacity and efficiency.}

\tikzstyle{my-box}=[
    rectangle,
    draw=black,
    rounded corners,
    text opacity=1,
    minimum height=1.5em,
    minimum width=5em,
    inner sep=2pt,
    align=center,
    fill opacity=.5,
    line width=0.5pt,
]
\tikzstyle{leaf}=[my-box, minimum height=1.5em,
    fill=green!3, text=black, align=left,font=\normalsize,
    inner xsep=2pt,
    inner ysep=5pt,
    line width=0.8pt,
]

\begin{figure}[!ht]
    \centering
    \resizebox{0.48\textwidth}{!}{ 
        \begin{forest}
        forked edges,
        for tree={
            grow=east,
            reversed=true,
            anchor=base west,
            parent anchor=east,
            child anchor=west,
            base=center,
            font=\large,
            rectangle,
            draw=black,
            rounded corners,
            align=left,
            text centered,
            minimum width=4em,
            edge+={black, line width=1pt},
            s sep=3pt,
            inner xsep=2pt,
            inner ysep=3pt,
            line width=0.8pt,
            ver/.style={rotate=90, child anchor=north, parent anchor=south, anchor=center},
        },
        where level=1{text width=8.2em,font=\small,}{},
        where level=2{text width=12em,font=\small,}{},
        [
        {Organizational \\ Decentralization}, fill=blue!5, font=\small
            [
            {Energy Efficiency \S \ref{sec:engergy_eff}}, fill=yellow!5
                [
                TowardDC~\cite{Wang2023TowardDC}{, }\\
                CAFE~\cite{Bian2023CAFECF}{, }\\
                LLMCarbon~\cite{Faiz2023LLMCarbonMT}{,}\\
                CAFTM~\cite{Park2024CarbonAwareAF}{,}\\
                CloudSimPer~\cite{jie2024cloudsim}{.}
                ,
                leaf, 
                text width= 11em, font=\small
                ]
            ]
            [
            {System Throughput \S \ref{sec:throughput}}, fill=yellow!5
                [
                Yugong~\cite{Huang2019YugongGD}{, }\\
                TFAS~\cite{Fan2024OnlineTF}{, }\\
                MAST~\cite{Choudhury2024MASTGS}{, }\\
                ACME~\cite{Hu2024CharacterizationOL}{. }\\
                ,
                leaf, 
                text width= 11em, font=\small
                ]
            ]
        ]
        \end{forest}
    }
    \caption{Taxonomy of related papers based on optimization objectives of organizational decentralization}
    \label{papers_organization}
\end{figure}

\subsection{Energy Efficiency}
\label{sec:engergy_eff}

Training LLMs within datacenters entails substantial energy consumption, making carbon efficiency a critical concern for large organizations. For a single job, predicting the carbon footprint of current training methods for large models can help optimize training strategies, thereby reducing carbon emissions \cite{Faiz2023LLMCarbonMT}. 
Resource selection across datacenters also plays a vital role, enabling a trade-off between model performance and environmental impact \cite{Bian2023CAFECF}
At the datacenter level, implementing job scheduling strategies that are aware of carbon emissions can further decrease the overall carbon footprint \cite{Park2024CarbonAwareAF,jie2024cloudsim}. 
In addition, digital twin technologies hold significant potential for enabling real-time monitoring, control, and optimization of datacenter operations, thereby enhancing energy efficiency during LLM training \cite{Wang2023TowardDC}.

\subsection{System Throughput}
\label{sec:throughput}

When training large models across datacenters, it is crucial to design resource scheduling policies that account for the characteristics of training processes, such as 4D parallelism and synchronization requirements, as well as the infrastructure characteristics including low-bandwidth inter-datacenter communication \cite{Fan2024OnlineTF,Hu2024CharacterizationOL}.
Traditional cross-datacenter schedulers often fall short, as they struggle to adapt to the dynamic and resource-intensive nature of LLM training workloads \cite{Huang2019YugongGD}. 
To address these challenges, MAST \cite{Choudhury2024MASTGS}, a global scheduling system, effectively orchestrates the training of LLama-3 \cite{Dubey2024TheL3} across 16,000 H100 GPUs, achieving both load balancing and fault tolerance across datacenters.

\paragraph{Discussion. }
Organizational decentralization is the preferred approach for large, well-resourced organizations to leverage massive computational resources for training ultra-large LLMs. While the global-scale distributed model training job scheduler MAST can coordinate tens of thousands of GPUs across multiple datacenters, the 16,000 GPUs used for training Llama-3 were confined to a single massive datacenter \cite{Dubey2024TheL3}.
Expanding such training frameworks to span multiple data centers in the future represents a pivotal pathway for developing ultra-scale LLMs.


\section{Case Study}

This section examines two representative models, Llama-3 and INTELLECT-1, which we considered as examples of organizational and community-driven decentralization, respectively. Comparative configurations are listed in Table \ref{tab:casestudy}.

\paragraph{Llama-3.}

Llama-3, an open-source foundation model family from Meta, scales up to 405 billion parameters \cite{Dubey2024TheL3}. 
It is trained on 15 T multilingual tokens, requiring a significant amount of FLOPs for computation.
To achieve this, Llama-3 utilizes 16,000 H100 GPUs with a global-scale scheduler \cite{Choudhury2024MASTGS} and 4D parallel training (as illustrated in Appendix \ref{sec:appendix_a}), achieving 38-43\% Model Fractional Utilization (MFU). 
Storage is managed by the Tectonic distributed file system \cite{Pan2021FacebooksTF}, offering 240 PB capacity and optimized throughput to reduce GPU idle time.
For networking, Llama-3 leverages RoCE and InfiniBand within a three-layer topology~\cite{BMGI2024,GangidiRDMAOE}, enhanced by load balancing and congestion control for efficient communication across 24,000 GPUs.
As a representative open-source LLM, Llama-3 leverages interconnected clusters orchestrated by a global-scale scheduler, providing critical insights for organizational decentralized LLM training paradigms.

\paragraph{INTELLECT–1.}

INTELLECT-1~\cite{Intellect1} , an open-source 10-billion-parameter LLM trained with decentralized resources. 
It employs hierarchical parameter aggregation and int8 quantization to minimize bandwidth usage, while VPN is integrated to ensure stability in low-bandwidth networks.
For fault tolerance, INTELLECT-1 utilizes ElasticDeviceMesh for node management and phased checkpointing to minimize training interruptions, with peer-to-peer transfer enabling rapid checkpoint recovery.
To optimize memory utilization, INTELLECT-1 integrates FSDP2~\cite{PyTorchFSDP} and CPU offloading, enhancing the system efficiency and scalability.
Furthermore, Intellect-2 \cite{Intellect2}, the successor model of INTELLECT-1, extends the capabilities of decentralized LLMs through decentralized post-training with reinforce learning.
This community-driven decentralized paradigm democratizes AI model development, preventing monopolization and fostering open-source innovation with decentralized resources.


\begin{table}[H]
    \centering
    \resizebox{0.49\textwidth}{!}{

\begin{tabular}{p{3.6cm}|p{4.3cm}p{4cm}} 
\toprule
{Model} & \textbf{Llama-3} & \textbf{INTELLECT-1} \\ 
\midrule

Parameter Scale & 405 B & 10 B \\ %
Resource Scale & 16K H100 GPUs & 112 H100 GPUs \\ 
Resource Distribution & Across pods, each with 3072 GPUs, in one datacenter & Across servers from 5 countries and 3 continents\\
Parallel Mechanism & 4D parallelism & Hybrid DiLoCo-FSDP2 \\
Training Time & 54 days & 42 days \\
Effective Training Time & $\geq 90\%$  & 83\% \\
Processed Tokens & 15 T & 1 T \\
\bottomrule
\end{tabular}

    }
    \caption{Comparison of training configurations of Llama-3 and INTELLECT-1}
    \label{tab:casestudy}
\end{table}

\section{Summary and Future Directions}

\subsection{Summary}

In this survey, we classify decentralized LLM training into two paradigms based on resource utilization: community-driven decentralization and organizational decentralization. By analyzing the characteristics of decentralized resources, reviewing relevant optimization methods from the literature, and examining two model cases, representing the applications of community-driven and organizational decentralization respectively, we provide a systematic overview of the current development landscape in decentralized LLM training.

\subsection{Future Directions}

We outline potential research directions spanning resource organization, model architecture, and training paradigms.

\paragraph{Scaling Law of Decentralized LLM Training.}

The scaling law for centralized LLM training primarily focus on computational power, data volume, and model size to optimize training strategies \cite{Hoffmann2022TrainingCL,Dubey2024TheL3}. 
In decentralized paradigms, however, the interplay between computational and network resources becomes significantly more complex, necessitating their inclusion in the scaling laws. 
For a given topology of computational resources with constrained bandwidth, there may exist a practical limit on scaling efficiency. Beyond this point, further increasing model size or local resources does not yield proportional improvements in global performance.
Exploring the scaling law for decentralized LLM training is essential for effectively coordinating global decentralized resources and enhancing their utilization efficiency.

\paragraph{Decentralized Resources Governance.}
Current efforts in decentralized LLM training primarily focus on resource utilization and management for individual model training processes.
As the community expands, effective governance of decentralized resources will become crucial for sustaining the development of decentralized LLM training. Challenges like communication bottlenecks, resource heterogeneity, and instability may stem from the resources themselves or arise from inefficient coordination at the resource abstraction layer of decentralized systems. 
Future research could prioritize developing mechanisms to optimize pricing and scheduling of decentralized resources in multi-tenant environments, thereby facilitating the proliferation of open-source models powered by decentralized infrastructure.

\paragraph{Training MLLM with Decentralized Resources.}

Decentralized training of multi-modal large language models (MLLMs) presents both significant opportunities and unique challenges compared to conventional LLM training. The inherent complexity of training MLLMs stems from different data types (e.g. text, images, audio), each requiring specialized processing modules. These heterogeneous modules complicate communication scheduling and resource allocation for distributed training \cite{distmm}.
When training MLLMs with decentralized resources, these heterogeneities can be further exacerbated.
Addressing these challenges through future research could unlock the full potential of decentralized multi-modal data, enabling scalable and efficient utilization of decentralized resources and significantly advancing the development of multi-modal AI systems.

\paragraph{Post-training with Decentralized Resources.}

Current research on utilizing decentralized resources for LLM training predominantly focuses on the pre-training stage. However, the post-training phase incorporating reinforcement learning (RL) is crucial for enhancing the reasoning capabilities of LLMs \cite{DeepSeekR1}. One distinctive characteristic of this phase lies in the intensive rollout generation \cite{Intellect2}, which is computation-intensive inference process without backward. Therefore, during RL-based post-training, decentralized methods should also optimize LLM inference serving on weaker nodes (e.g., serving LLMs with frameworks like vLLM \cite{pagedAttn} or SGLang \cite{sglang} on consumer GPUs, which is non-trivial), while pre-training can only consider training period optimization.
Future research could focus on the joint optimization of inference rollout and backward during LLM post-training with decentralized resources, which can expand the capability boundaries of decentralized LLMs, thereby enhancing accessibility to LLM services for broader communities.

\section*{Limitations}

This survey focuses on decentralized LLM training from a resource-driven perspective, but several limitations should be noted. First, important issues such as data distribution and privacy protection are not covered, as they diverge from the scope of our survey. Second, given the rapid advancements in LLM development, some recent developments may have been inadvertently overlooked despite our efforts to include more relevant research. 

This paper is a survey and does not involve the development of new artifacts or data collection. Therefore, it poses no direct potential risks.

\section*{Acknowledgment}

This work is supported in part by National Key Research and Development Project of China (Grant No.~2023YFF0905502), National Natural Science Foundation of China(Grant No.~92467204 and 62472249), Shenzhen Science and Technology Program (Grant No.~JCYJ20220818101014030 and~KJZD20240903102300001) and Natural Science Foundation of Top Talent of SZTU(Grant No.~GDRC202413). 
We thank the anonymous reviewers for their efforts, which have helped improve the quality of this paper.

\bibliography{custom}
\appendix
\tikzstyle{my-box}=[
    rectangle,
    draw=black,
    rounded corners,
    text opacity=1,
    minimum height=1.5em,
    minimum width=5em,
    inner sep=2pt,
    align=center,
    fill opacity=.5,
    line width=0.5pt,
]
\tikzstyle{leaf}=[my-box, minimum height=1.5em,
    fill=green!3, text=black, align=left,font=\normalsize,
    inner xsep=2pt,
    inner ysep=5pt,
    line width=0.8pt,
]

\begin{figure*}[!ht]
    \centering
    \resizebox{0.98\textwidth}{!}{ 
        \begin{forest}
        forked edges,
        for tree={
            grow=east,
            reversed=true,
            anchor=base west,
            parent anchor=east,
            child anchor=west,
            base=center,
            font=\small,
            rectangle,
            draw=black,
            rounded corners,
            align=left,
            text centered,
            minimum width=4em,
            edge+={black, line width=1pt},
            s sep=3pt,
            inner xsep=2pt,
            inner ysep=3pt,
            line width=0.8pt,
            ver/.style={rotate=90, child anchor=north, parent anchor=south, anchor=center},
        },
        where level=1{text width=8em,font=\small,}{},
        where level=2{text width=12em,font=\small,}{},
            [
            {Community-driven Decentralization}, ver,
            fill=blue!5,
                [
                {Communication \S \ref{sec:comm_optimization}}, fill=yellow!5
                    [
                    \textbf{Reduce Communication Frequency}{ }\\
                    Gaia~\cite{Hsieh2017GaiaGM}{, }
                    Co-learning~\cite{Xu2018CollaborativeDL}{, }\\
                    DeDLOC~\cite{Diskin2021DistributedDL}{, }
                    Varuna~\cite{Athlur2022VarunaSL}{, }
                    DiLoCo~\cite{Douillard2023DiLoCoDL}{, }\\
                    DTFM~\cite{Yuan2022DecentralizedTO}{, }
                    SWARM~\cite{Ryabinin2023SWARMPT}{.} \\
                    \textbf{Reduce Communication Intensity}{ }\\
                    Moshpit~\cite{Ryabinin2021MoshpitSC}{, }
                    AQ-SGD~\cite{Wang2022FinetuningLM}{, }\\
                    DTFM~\cite{Yuan2022DecentralizedTO}{, }
                    SWARM~\cite{Ryabinin2023SWARMPT}{, }
                    Ravnest~\cite{Menon2024RavnestDA}{, }\\
                    OpenDiLoCo~\cite{jaghouar2024opendiloco}{, }
                    StellaTrain~\cite{StellaTrain2024}{, }
                    Petals \cite{Borzunov2022PetalsCI}{.}
                    ,leaf, 
                    text width= 36.5em, font=\small
                    ]
                ]
                [
                {Heterogeneity \S \ref{sec:hete_aware}}, fill=yellow!5
                    [
                    \textbf{Network Heterogeneity}{. }\\
                    BACombo~\cite{Jiang2020BAComboBandwidthAwareDF}{, }
                    NETSTORM~\cite{Li2024AcceleratingGM}{, }
                    SAGQ~\cite{Fan2023SelfAdaptiveGQ}{, }\\
                    DeDLOC~\cite{Diskin2021DistributedDL}{, }
                    Holmes~\cite{Yang2023HolmesTD}{, }
                    DTFM \cite{Yuan2022DecentralizedTO}{.}\\
                    \textbf{Device Heterogeneity}{ }\\
                    Learning@home~\cite{Ryabinin2020TowardsCT}{, }
                    Dlion~\cite{Hong2020DLionDD}{, }
                    ATOM \cite{Wu2024ATOMAT}{, }\\
                    Petals~\cite{Borzunov2022PetalsCI}{, }
                    SWARM~\cite{Ryabinin2023SWARMPT}{, }
                    Ravnest~\cite{Menon2024RavnestDA}{. } \\
                    ,                    
                    leaf, 
                    text width= 36.5em, font=\small
                    ]
                ]
                [
                {Fault Tolerance \S \ref{sec:fault_tolerance}}, fill=yellow!5
                    [
                    DeepSpotCloud~\cite{Lee2017DeepSpotCloudLC}{, }
                    DDNN~\cite{Teerapittayanon2017DistributedDN}{, }
                    ATOM~\cite{Wu2024ATOMAT}{, }\\
                    DeDLOC~\cite{Diskin2021DistributedDL}{, }
                    Petals~\cite{Borzunov2022PetalsCI}{, }
                    SWARM~\cite{Ryabinin2023SWARMPT}{, }\\
                    FusionAI~\cite{Tang2023FusionAIDT}{, }
                    L@H~\cite{Ryabinin2020TowardsCT}{, }
                    OpenDiLoCo~\cite{jaghouar2024opendiloco}{, }\\
                    DiLoCo~\cite{Douillard2023DiLoCoDL}{, }
                    Position~\cite{Lu2024PositionET}{, }
                    Moshpit \cite{Ryabinin2021MoshpitSC}{. }
                    ,
                    leaf, 
                    text width= 36.5em, font=\small
                    ]
                ]
                [
                {Economics \S \ref{sec:economics}}, fill=yellow!5
                    [
                    DeepSpotCloud~\cite{Lee2017DeepSpotCloudLC}{, }
                    Varuna~\cite{Athlur2022VarunaSL}{, }
                    FusionAI~\cite{Tang2023FusionAIDT}{, }\\
                    MLTC~\cite{Strati2024MLTW}{, }
                    HowCW~\cite{Isenko2023HowCW}{, }
                    TowardsGT~\cite{Phalak2024TowardsGT}{, }\\
                    StellaTrain~\cite{StellaTrain2024}{.}
                    ,
                    leaf, 
                    text width= 36.5em, font=\small
                    ]
                ]
            ]
        \end{forest}
    }
    \caption{Taxonomy of related papers based on optimization objectives of community-driven decentralization.}
    \label{fig:papers_community}
\end{figure*}

\begin{table*}[ht]
\centering

\resizebox{\linewidth}{!}{%
\begin{tabular}{lllll}
\hline
\textbf{Parallel Strategy} & \textbf{Data Parallelism} & \textbf{Pipeline Parallelism} & \textbf{Tensor Parallelism} & \textbf{Context Parallelism} \\
\hline
Parallel Granularity & Data batches & Model stages and data batches & Intra-layer tensor slices & Partitioned sequences \\
Model Partition & Full model & Model block partitioned by stage & Model block partitioned by tensor slice & Full model \\
Communication & All-reduce full gradients & Inter-stage activations and gradients & All-reduce/All-gather intra-layer states & All-to-all attention KV tensors\\
Memory Usage & Full model duplication & Sharded model and activations & Sharded model and activations & Sharded KV cache and activations\\
Scalability & Good for large data batches & Good for inter-node communication & Good for intra-node communication & Good for long sequence processing \\
Example & PyTorch DDP \cite{pytorchDDP} & GPipe \cite{GPipe2019huang} & Megatron-LM \cite{Narayanan2021EfficientLL} & DeepSpeed-Ulysses \cite{Ulysses} \\
\hline
\end{tabular}}%

\caption{Comparison of four primary types of parallelism in LLM training.}
\label{tab:parallel-comparison}
\end{table*}

\section{Parallel Strategies in LLM Training}
\label{sec:appendix_a}

Training LLMs demands substantial computational resources and advanced parallel strategies to achieve efficiency. To address the challenges posed by the massive scale of LLMs, researchers have developed multi-dimensional parallel strategies, including Data Parallelism (DP), Tensor Parallelism (TP), Pipeline Parallelism (PP), and Context Parallelism (CP). Among these, TP and PP can be regarded as specialized forms of Model Parallelism (MP). Context Parallelism (CP) \cite{MegatronCore2023,Ulysses} has emerged as a complementary strategy, which operates at the token level by slicing input sequences across devices, enabling scalable and efficient training of long-context LLMs. DP and PP strategies have been implemented within decentralized LLM training.
When these techniques are strategically combined during the training of Llama-3 \cite{Dubey2024TheL3}, they collectively enhance throughput, reduce memory footprint, and optimize resource utilization.
A detailed comparison of these four parallelism strategies is presented in Table \ref{tab:parallel-comparison}. 

\section {Literature Summary}
\label{sec:appendix_b}

More comprehensive studies associated with community-driven decentralization
are presented in Figure \ref{fig:papers_community}. 
While this survey encompasses a broad spectrum of research, our main text primarily focuses on works that specifically target LLMs, ensuring a more in-depth and focused analysis.

\section{Distributed Hash Tables (DHTs)}
\label{sec:appendix_c}

DHTs are a class of decentralized storage systems designed to provide scalable, fault-tolerant, and efficient key-value lookups across a large set of networked nodes. Unlike traditional centralized hash tables, where a single server manages all mappings between keys and values, DHTs distribute this responsibility across multiple peers, each responsible for a subset of the key space.
These features make DHTs an enabling infrastructure for robust and elastic decentralized LLM training systems, especially under environments with high node failure ratios and heterogeneous conditions.

First, DHTs can facilitate recovery from failures by storing training states \cite{Isenko2023HowCW}. In Learning@home~\cite{Ryabinin2020TowardsCT}, expert checkpoints are stored in a DHT, allowing newly joined nodes to retrieve the latest state of failed ones and resume training seamlessly. Similarly, ATOM~\cite{Wu2024ATOMAT} leverages DHTs for asynchronous training, enabling task reallocation and training states recover.

Second, DHTs can act as metadata stores to coordinate task redistribution and enable resilient system reconfiguration. For instance, Petals~\cite{Borzunov2022PetalsCI} uses DHTs to manage model shard placement, allowing the system to rebalance and recover from failures during collaborative fine-tuning. SWARM~\cite{Ryabinin2023SWARMPT} extends this by integrating stochastic pipeline rewiring, allowing the PP process to filter failed nodes and redistribute workloads by iteration. FusionAI~\cite{Tang2023FusionAIDT} adopts a similar design by combining metadata with an agent to handle task reassignment and node recovery.

Additionally, DHTs enable robust coordination of decentralized communication and update mechanisms, even under high node volatility. Moshpit~\cite{Ryabinin2021MoshpitSC} uses DHTs to dynamically form groups for gradient averaging, ensuring that partial updates are aggregated reliably despite frequent node failures.
These approaches collectively highlight that DHTs can enhance the fault tolerance during decentralized LLM training.

\end{document}